\documentstyle[12pt]{article}
\begin{document}
\begin{center}
{\ {\Large {\bf Coulomb Final State Interaction }}}\\
\bigskip
{\ {\Large {\bf in Pion Interferometry }}}  \\
\bigskip
{\ {\Large {\bf for the Processes of High Multiplicity }}} \\

\vskip 0.8cm

{\large {\sl D.V.~Anchishkin$^{*,1}$, W.A.~Zajc$^{**}$, G.M.~Zinovjev$^{***}$}}
\\
\vskip 0.80cm
$^{*}$University of Jyv\"{a}skyl\"{a}, Department of Physics, \\
P.O.Box 35, FIN-40351 Jyv\"{a}skyl\"{a}, FINLAND\\
E-mail: ANCHISHKIN@jyfl.jyu.fi\\
\vskip 0.40cm
$^{**}$Nevis Laboratories,\\ Columbia University, \\
Irvington, NY 10533, USA\\
E-mail: ZAJC@nevis1.nevis.columbia.edu\\
\vskip 0.40cm
$^{***}$Bogolyubov Institute for Theoretical Physics, \\
 National Academy of Sciences of Ukraine, \\
252143 Kiev-143, UKRAINE \\
E-mail: GEZIN@gluk.apc.org\\

\end{center}

\vskip 1.2cm

\begin{abstract}

The corrections for two pion correlations due to electromagnetic
final state interactions at high secondary multiplicities are
investigated. It is shown that these result in a noticeable deviation
from the standard Gamov factor.
This conclusion changes drastically in a model of the pion system with
expansion.
The critical  parameter which determines the size of these effects
is found to be the ratio of the relative velocity of detected pions to
the velocity of the pair center-of-mass (in the fireball rest frame). In
particular, when this parameter is much less than unity the pion pair escapes
the initial high density region promptly
and the distortion of the mutual Coulomb
potential is weak.

\end{abstract}

\vspace{25mm}

\noindent
----------------------------------------------------------------------------\\
\noindent
$^{1}$Permanent address:
Bogolyubov Institute for Theoretical Physics,
National Academy of Sciences of Ukraine, 252143 Kiev-143, UKRAINE
(E-mail: ANCH@gluk.apc.org)

\pagebreak

\vskip 0.5cm

The fundamental observable for intensity interferometry in hadron
physics is the relative momentum spectrum between two identical particles.
For two like-charged pions, the final-state Coulomb interaction
modifications to this spectrum result in a correction which has
typically been considered to be tractable and relatively accurate.
This expectation is based on the significantly different length
scales between the strong  ($\propto 1/m_\pi $) and
the Coulomb ($\propto 1/m_\pi \alpha$) interactions in the problem
\cite{Sakharov,bf} (here $m_\pi$
is a pion mass and $\alpha $ is a fine structure constant).
In this case the the correction may be
treated with the Schr\"{o}dinger equation, resulting in the well-known
Gamov factor $G({\bf p}_{1},{\bf p}_{2})$. The nominal quantity
expressing the correlation function in terms of experimental
distributions \cite{boal}
\begin{equation}
C({\bf p} _{1},{\bf p} _{2})=\frac{<n>^{2}}{<n(n-1)>} \frac{\displaystyle \frac
{d^{6}\sigma }{d^{3}p_{1}d^{3}p_{2}} }
{\displaystyle \frac{d^{3}\sigma }{d^{3}p_{1}} \frac{d^{3}\sigma }
{d^{3}p_{2}} } ,
\label{eq1}
\end{equation}
\noindent where $<n>$ is the particle multiplicity, $d^{3}\sigma /d^{3}p$
and $d^{6}\sigma /d^{3}p_{1}d^{3}p_{2}$ are the single-pion and two-pion
cross sections,
is then given (due to the factorization of the corresponding matrix element)
in terms of the product of a Gamow factor
multiplying the model correlations
\cite{GKW}
\begin{equation}
C({\bf p} _{1},{\bf p} _{2})=G({\bf p} _{1},{\bf p} _{2})
C_{model}({\bf p} _{1},{\bf p} _{2}) .
\label{eq6.0}
\end{equation}

The standard derivation of the Gamov factor \cite{Messiah}, obtained
from the solution $\psi (\bf r)$ of the nonrelativistic Schr\"{o}dinger
equation with the Coulomb potential, leads to

\begin{equation}
G(Q)=\mid \psi (0)\mid ^2= \frac{2\pi \eta }{e^{2\pi \eta }-1} ,
\label{eq10}
\end{equation}

\noindent where

\begin{equation}
\eta = \eta _{0} \equiv \frac{\alpha m_{\pi }}{Q} ,
\label{eq9}
\end{equation}

\noindent where $Q$ is the relative  momentum between the particles.
It should be
mentioned that, in terms of the pion momentum $k$ in the pair center-of-mass,
the relative
pion momentum $Q=2k$ coincides with invariant relative momentum
$Q_{inv}\equiv [(k_{1}+k_{2})^{2}-4m_{\pi }^{2}]^{1/2}$
where $k_{1}$ and $k_2$ are pion four-momenta in an arbitrary frame.
In order to better illustrate the more complicated cases discussed below,
it is worthwhile to reconsider the calculation of
this phenomenon as a quantum mechanical tunneling process (see for instance
\cite{anch}).

In the quasi-classical approximation the probability for a pion of mass
$m_\pi$  starting
from the point $r_{2}$ to reach the distance $r_{1}$ under the barrier is
related to the quantity

\begin{equation}
\eta _{pen}(r_{1},r_{2})\equiv \frac{1}{\pi } \int ^{r_2}_{r_1} q(r)dr ,
\label{eq12.1}
\end{equation}

\noindent with

\begin{equation}
q(r)=\left[ 2m_{\pi }\left( V(r)-E_{kin}^0\right) \right] ^{1/2} ,
\label{eq13}
\end{equation}

\noindent being the so-called ``subbarrier quasi-momentum".
The equation
$V(r_2)=E_{kin}^{0}$ determines the classical turning point.

For the pure Coulomb barrier $V_{Coul}=\alpha /2r$
the calculation of Eq.~(\ref{eq12.1}) gives

\begin{equation}
\eta _{pen}(r_{1},r_{2})=\frac{1}{\pi } \left[ I(r_2)-I(r_1)
\right] ,
\label{etnr}
\end{equation}

\noindent with

\begin{equation}
I(r)=rq(r)-\eta _{0}\arcsin \left[ 1-2\frac{E_{kin}^{0}}{V_{Coul}(r)}
\right] ,
\label{inr}
\end{equation}

\noindent where the factor of $1/2$ in the potential accounts for
the  usage of the
physical pion mass, not the reduced one, in Eq.~(\ref{eq13}).
For  penetration to $r_{1}=0$, corresponding to production of
the two pions at the same space-time point,
one obtains

\begin{equation}
\eta _{pen}(r_{1}=0,r_{2})=\frac{\alpha m_{\pi }}{Q}=\eta _{0} ,
\label{eq14}
\end{equation}

\noindent identical with (\ref{eq9}).

Turning to the high multiplicity case,
the relation between the  two-particle
electromagnetic potential and the local charge density is given by

\begin{equation}
 \nabla^2 \phi ({\bf r})=-4\pi e (n^{(+)}-n^{(-)})\ ,
\label{15}
\end{equation}

\noindent where $e$ is the elementary charge and
and where the density of charged mesons $n^{(\pm )}$
is then related  to  that of neutral mesons $n^{(0)}$
via a Boltzmann factor:
\begin{equation}
n^{(\pm )}=n^{(0)}\ exp\left(\mp \frac{{\it e}\phi }{T}\right) \ ,
\label{16}
\end{equation}
Here $n^{(0)}$ is the density of $\pi ^{0}$-mesons at
the freeze-out temperature $T$,
which coincides with the equilibrium density of charged pions
in the absence of  the Coulomb interaction (we consider symmetrical nuclear
matter).
When $ e\phi \ll T$ the Eq.(\ref{16}) can be rewritten as

\begin{equation}
n^{(\pm )}=n^{(0)}\, \left( 1 \mp \frac{{\it e}\phi }{T}\right)\ ,
\label{17}
\end{equation}

\noindent (this requires that the pions are not closer than $\sim10^{-2}$~fm to
one another at $T\approx 200\ MeV$) so that

\begin{equation}
\nabla^2 \phi ({\bf r})=\frac{4\pi  e^2}{T} (    2n^{(0)}   )
\phi ({\bf r})\ ,
\label{18}
\end{equation}

\noindent where $e^{2}=\alpha $.

We can write down immediately the well-known solution of the Eq.(\ref{18}) as
the screened Coulomb potential

\begin{equation}
\phi _{\pi ^{\pm }}(r)=\pm {\it e}\frac{e^{-r/R_{scr}}}{r}\ ,
\label{19}
\end{equation}

\noindent where

\begin{equation}
\frac{1}{R_{scr}} =
\sqrt{\frac{8\pi }{3} \alpha } \cdot \sqrt{\frac{n_{\pi }}{T} }\ ,
\label{20}
\end{equation}

\noindent with potential energy $U_{\pi \pi }=\alpha \exp(-r/R_{scr})/r $
for the like-sign pions.

To evaluate the correction to the Gamow factor, we will use this screened
Coulomb potential to re-calculate the penetration parameter given by
Eq.(\ref{eq12.1}). This in turn  requires  an estimate
of the pion density and freeze-out temperature.
We will calculate this for the extreme case, i.e.,
just after freeze-out, when pions occupy the
volume \cite{sinyukov} (we take the same temperature $T_{f}=190\ MeV$)

\begin{equation}
V_{f}=\pi R_{f}^{2}2\tau _{f}\sinh{\frac{\Delta y}{2} }\ ,
\label{23}
\end{equation}

\noindent where for Pb-Pb collisions $R_{f}=R_{Pb}\approx 7.1\ fm$,
$\tau _{f}=10\ fm$ and $\Delta y \approx 6$.
For multiplicities we assume
\cite{satz} : {\it LHC}:~$N_{\pi }=8000$, {\it SPS}:~$N_{\pi }=800$, which
gives for screening radii
{\it LHC}:~$R_{scr}=7.9\ fm$, {\it SPS}:~$R_{scr}=25\ fm$ .
The latter looks rather controversial at first sight (though for a
pure Coulomb interaction $R_{scr} = \infty $). Indeed, using Eq.~(\ref{23})
for the unit interval of rapidity (one might consider that more natural,
in a sense)
we obtain much smaller value of corresponding screening radius. None the
less we cite this quantity here as an upper bound in existing experimental
conditions.

The results of these calculations together with the standard Gamov factor
are plotted in Fig.~1.
Also shown there is a curve obtained by direct substitution of the
experimental values obtained by NA44 \cite{atherton} into
Eq.~(\ref{23}), with
$T_{f}=187\ MeV$, $\tau _{f}=R_{L}\approx 6.0 \ fm$, $R_{T}=6\ fm$,
$\Delta y=3$ and  $dN/dy=40$.
These parameters produce an even smaller screening radius
($R_{scr}=19.3\ fm$) than our nominal SPS value (and hence a larger
correction to the standard Gamov factor),
which follows from the
smaller values of
$R_{scr}\approx 4\sqrt{T_{f}/n_{\pi } } $ at the
NA44 experimental conditions.
Based on these consideration alone, it would appear
that a substantial correction to the Gamow factor
is already required by the existing experimental data.
This is particularly true for kaon
interferometry since the pion medium screens the K-K Coulomb final
state interaction as well, and the length scale for the K-K Gamov
factor is larger than that for pions by the ratio of their masses.

However, it is important to note that correction factors presented
in Fig.~1 are an upper bound, in that they do not incorporate
the expansion of the pion source after freeze-out.
We next turn to a more realistic calculation explicitly incorporating
expansion.
It is
intuitively clear that the correction factor for the fast pairs will
approach the standard Gamov factor and for the very slow ones the
estimate obtained should be valid.

\bigskip

We parametrize the spherical expansion of the pion source

\begin{equation}
n(R)=n_{f}\frac{R_{f}^{2}}{R^{2}}\ ,
\label{24}
\end{equation}

\noindent in terms of the parameters $n_f$, the freeze-out pion density;
the corresponding radius $R_{f}$
and the freeze-out temperature $T_{f}$.
In this model the spatial volume of the expanding pion system
in the solid angle
$\Omega $ increases as $\Omega \cdot R^{2}$, where  $R$ is the distance
from the center of the fireball.  Then, the corresponding potential
is the solution of the Maxwell equation

\begin{equation}
\left( \nabla^2 - \partial _{t}^{2}\right) \, \phi ({\bf r})=
\frac{8\pi e^2 }{3\, T} n_\pi \, \phi ({\bf r})\ ,
\label{51}
\end{equation}

\noindent where we put $n^{(0)} \approx n_{\pi }/3$ as before.  Now
introducing the distance $r$ between two detected particles we
have

\begin{equation}
R \approx v_{cm}\cdot t \ , \hspace{1.5cm} r \approx v_{rel}\cdot t \ ,
\label{26}
\end{equation}

\noindent where $v_{cm}$ is the velocity of the two-particle center-of-mass
in the fireball rest frame and $v_{rel}$ is the relative velocity of the
particles ($v_{rel}=Q/m$, $t=0$ is fixed on the freeze-out hyper-surface).
Then

\begin{equation}
R=\frac{v_{cm}}{v_{rel}} \cdot r\ .
\label{27}
\end{equation}

\noindent so that Eq.(\ref{51}) may be rewritten as

\begin{equation}
\left( \nabla^2 - v_{rel}^{2} \partial _{r}^{2}\right) \, \phi ({\bf r}) =
\frac{c^{2}(Q)}{r^{2}} \, \phi ({\bf r})\ ,
\label{52}
\end{equation}

\noindent with $c^{2}(Q)$ defined as

\begin{equation}
c^{2}(Q) = \frac{8\pi e^{2}}{3} \frac{R_{f}^{2}n_{\pi }}{T}
\frac{v_{rel}^{2}}{v_{cm}^{2}}\ .
\label{53}
\end{equation}

\noindent Fixing the screening radius at the freeze-out temperature and
density as

\begin{equation}
R_{scr}^{f}=
\sqrt{\frac{3T_{f}}{8\pi \alpha n_{f}} } \ ,
\label{53a}
\end{equation}

\noindent we have

\begin{equation}
c(Q) =  \frac{R_{f}}{R_{scr}^{f}} \frac{v_{rel}}{v_{cm}} \ .
\label{53b}
\end{equation}

\noindent and Eq.(\ref{51}) can be rewritten as

\begin{equation}
r^{2}(1-v_{rel}^{2}) \frac{d^{2}\phi }{dr^{2}} + 2r\frac{d\phi }{dr} -
c^{2}(Q)\, \phi \, =\, 0 \ .
\label{54}
\end{equation}

\noindent This equation has the following solution

\begin{equation}
\phi = \frac{eR_{0}^{a-1}}{r^{a}}\ ,
\label{55}
\end{equation}

\noindent where $R_{0}$ is a scale parameter to fix the dimensions of
the potential $\phi $. The exponent $a$ is a solution of the quadratic
equation resulting from the substitution of the ansatz (\ref{55}) into
Eq.(\ref{52}) and takes the form (assuming proper asymptotic behaviour of
the potential $\phi $)

\begin{equation}
a = \frac{1}{2} \left[ \frac{1+v_{rel}^{2}}{1-v_{rel}^{2}} +
\sqrt{\left(\frac{1+v_{rel}^{2}}{1-v_{rel}^{2}} \right)^{2} +
\frac{4c^{2}(Q)} {1-v_{rel}^{2}} } \, \right] \ ,
\label{56}
\end{equation}

\noindent For small pion relative velocity $v_{rel}$ (corresponding to
$Q  \sim 0-30$~MeV) the above equation reduces to the
simple expression

\begin{equation}
a=\frac{1}{2} +\frac{1}{2}\sqrt{1+4c^{2}(Q)} \ .
\label{56a}
\end{equation}

\noindent Both are shown for several sets of parameters in Fig.~2.
In the interval of interest ``$a$'',
and hence deviations from a pure
Coulomb field, increases with increasing relative velocity.
This intriguing behavior is directly related to the
``Hubble-like'' expansion implied by Eq.~(\ref{27}),
and becomes quite understandable if we remember
that the pion density decreases
(hence $R_{scr} \rightarrow \infty $)  with $R$ increasing.
The expansion thus results in modifications to the
Coulomb potential that are of power-law,
not exponential form, in contrast to that static result given by
Eq.~(\ref{19}). We can treat
the corrected potential obtained in Eq.(\ref{55}) in terms of an
effective charge distribution

\begin{equation}
e_{eff}=e\left(\frac{R_{0}}{r} \right)^{a-1}\ ,
\label{58}
\end{equation}

\noindent which we are going to average over using the following procedure.
If we confine out attention to $v_{rel} \ll 1$,
Eq.(\ref{52}) can be rewritten in a Poisson-like form
$ (\nabla^2 - \kappa ^{2})\phi ({\bf r})=0$, with $\kappa \equiv c(Q)/r $,
which is equivalent to a $r$-dependent screening radius, i.e.,
\begin{equation}
R_{scr}(r)=\frac{r}{c(Q)} \ .
\label{25}
\end{equation}
Since, as shown in Fig.~2, the deviation of the
potential (\ref{55}) from the pure
Coulomb form in the region of small relative pion momenta $Q\le 30\ MeV$
is small,
we first ignore the $r$ dependence of $\kappa $ to
obtain the solution for the electromagnetic $\pi \pi $ potential in the
form of Eq.~(\ref{19}),
then substitute the $r$-dependent $\kappa $ into Eq.~(\ref{19}) to find
\begin{equation}
U_{\pi \pi } = \frac{\alpha e^{-c(Q)}}{r}\ ,
\label{59}
\end{equation}

\noindent
Thus, there is no an exponential dependence of $U_{\pi \pi }$ on $r$.
Instead,
the numerator on r.h.s. of Eq.(\ref{59}) represents the averaged charge
distribution (\ref{58}) squared
(we are now considering the potential energy $U_{\pi\pi} $ rather than the
 electric potential $\phi$, hence the extra factor of charge leading
 the $\alpha$).

It is clear from Eq.~(\ref{53b}) that when the ratio
$v_{rel}/v_{cm} \ll 1$ ($R_{f}$ and $R_{scr}^{f}$
are of the same order for high multiplicities) the renormalized
constant $\alpha exp[-c(Q)]$ is close to the bare value of
$\alpha $. Moreover, the same qualitative result comes from
the $r$-behaviour of the screening radius (\ref{25}) when it approaches the
asymptotic value $R_{scr}=\infty $ (Coulomb law)
with increasing $r$.
The quantity $c(Q)$ increases with
relative pion momentum leading to larger
deviations from the Coulomb
potential and agreement with the features of the
potential (\ref{55}) (see discussion after Eq.~(\ref{56a})).

Based on these considerations,
solving the Schr\"{o}dinger equation for the screened potential gives results
similar to Eqs.~(\ref{eq10}) and (\ref{eq9}),  but with the renormalized
$\alpha $, so that
\begin{equation}
\eta =\frac{\alpha m_{\pi }}{Q} exp[-c(Q)]=
\eta _{0}\cdot exp\left[ -\frac{v_{rel}}{v_{cm}} \cdot
\frac{R_{f}}{R_{scr}^{f}} \right] \ ,
\label{33}
\end{equation}

\noindent and taking into account the Eq.~(\ref{20}) we have

\begin{equation}
\frac{R_{f}}{R_{scr}^{f}}  =
\sqrt{2\alpha \frac{N_{\pi }}{R_{f}T_{f}} } \ .
\label{35}
\end{equation}

This result shows explicitly that the large modifications to the Gamow function
from to screening are weakened in the expansion model
due to the small value of $v_{rel}/v_{cm}$.
Specific examples of this are shown in Figs.~3 and 4.

Since for much of the parameter space
$c(Q)\ll 1$, it is useful to further approximate $\exp[-c(Q)] $.
Again referring to the NA44 data \cite{atherton}, and considering
the direction transverse to the collision axis where
$v_{cm}\approx p_{T}/m_{T}$ for the set of parameters

\begin{equation}
p_{T}\geq 150\ MeV/c,\ \ N_{\pi }\leq 200,\ \ R_{f}\geq 4\ fm,\ \ T_{f}\geq
180\ MeV,\ \ Q\leq 40\ MeV/c,
\label{36}
\end{equation}

\noindent
one has $\exp(-c(Q)) \simeq 1- c(Q)$, and the screening of the Coulomb
interaction reduces to shift $\eta _{0}$

\begin{equation}
\eta \simeq \eta _{0} - \Delta \eta \ ,
\label{37}
\end{equation}

\noindent with

\begin{equation}
\Delta \eta = \alpha \frac{m_{T}}{p_{T}} \cdot
\sqrt{2\alpha \frac{N_{\pi }}{T_{f}R_{f}} }\ ,
\label{38}
\end{equation}

\noindent
being independent of relative pion momentum $Q$.
(Clearly, for high
multiplicities and for small $p_{T}$ this approximation will be violated.)

For the case of cylindrical geometry we have

\begin{equation}
n_{f}=\frac{N_{\pi }}{\pi R_{f}^{2}2\tau _{f}\sinh{\frac{1}{2} \Delta y} }\ .
\label{35.1}
\end{equation}

\noindent and for the ratio (\ref{35})

\begin{equation}
\frac{R_{f}}{R_{scr}^{f}}  =
2\sqrt{\frac{\alpha }{3} \frac{N_{\pi }}
{T_{f}\tau _{f}\sinh{\frac{1}{2} \Delta y} } } \ ,
\label{35.2}
\end{equation}

\noindent and finally

\begin{equation}
\eta = \eta _{0}\cdot \exp\left[ -\frac{2Qm_{T}}{p_{T}m_{\pi }}
\sqrt{\frac{\alpha N_{\pi }}{3T_{f}\tau _{f}\sinh{\frac{1}{2} \Delta y}
} } \
\right] \ .
\label{35.3}
\end{equation}

Again, for the set of parameters where one might expect this geometry
to apply,

\begin{equation}
p_{T}\geq 150\ MeV/c,\ \ N_{\pi }\leq 2000,\ \ \tau _{f}\geq 10\ fm,\ \
\Delta y\geq 3,\ \ T_{f}\geq 180\ MeV,\ \ Q\leq 40\ MeV/c\ .
\label{39}
\end{equation}

\noindent we obtain the shift $\Delta \eta $ as

\begin{equation}
\Delta \eta = 2\alpha \frac{m_{T}}{p_{T}} \cdot
\sqrt{\frac{\alpha N_{\pi }}{3T_{f}\tau _{f}\sinh{\frac{1}{2} \Delta y}
} }\ ,
\label{40}
\end{equation}

\noindent which is also independent of relative pion momentum.

Fig.~3 shows the role of screening for the correction factor $G_{scr}$
behaviour evaluated at several sets of parameters together with the
standard Gamov factor $G_{0}$. Their ratio $G_{0}/G_{scr}$ plotted in
Figs.~4 and 5 show that correlators measured  in the
transverse direction develop a well-pronounced dependence on the transverse
momentum of the pair in the region of $p_{T}\leq 150\ MeV/c$ \cite{beker}
where the deviation from standard Gamov factor behaviour increases.

Our considerations show that for future LHC and RHIC experiments the
screening radius of Coulomb interaction at the freeze-out density and
temperature could be comparable with the source size and therefore the
factorization of Eq.~(\ref{eq6.0}) \cite{GKW} is no longer valid.
However, this conclusion
changes drastically with the inclusion
(switching on)
of  expansion for the pion system.
We would like to emphasize that this main result could be model
independent. The detailed evaluation of the Gamow correction in pion
interferometry analysis at very high multiplicities of secondary particles
in the picture of an expanding  fireball reveals an important regulating
parameter what is the ratio of relative velocity of the detected pions
and their center-of-mass velocity in the rest frame of a fireball
$v_{rel}/v_{cm}$.

\bigskip

{\bf Acknowledgements:}
We would like to acknowledge useful conversations with
D.~Ferenc, G.~Leksin,
B.~L\"{o}rstad, V.~Lyuboshitz, L.~McLerran, the late M.~Podgoretsky,
V.~Ruuskanen, H.~Satz, J.~Schukraft and A.~Stavinsky.
One of us (D.~A.) wishes to thank his colleagues at Physics Department
of the University of Jyv\"{a}skyl\"{a} for their warm hospitality.
This  work was supported
by INTAS Grant No.94-3941 and DOE FG02-86-ER40281.

\bigskip

\pagebreak


\begin{thebibliography}{18}
\vspace{0.3cm}

\bibitem{Sakharov}
A.D.Sakharov,
 Interaction of electron and positron during pair creation,
  Soviet JETP {\bf 18} (1948) 631-635.

\bibitem{bf}
V.N.Baier, V.S.Fadin,
 Coulomb final state interaction,
  Soviet JETP {\bf 57} (1969) 225-231, No.1(7).

\bibitem{boal}
D.H.Boal, C.-K.Gelbke and B.K.Jennings,
 Intensity interferometry in subatomic physics,
  Rev.\ Mod.\ Phys.\ {\bf 62} (1990) 553-602.

\bibitem{GKW}
M.Gyulassy, S.K.Kauffmann and L.W.Wilson,
 Pion interferometry of nuclear collisions. I. Theory,
  Phys.\ Rev.\ {\bf C20} (1979) 2267-2292.

\bibitem{Messiah}
A.Messiah,
 Quantum Mechanics Vol.1,
  North-Holland, Amsterdam 1961.

\bibitem{anch}
D.Anchishkin, G.Zinovjev,
 Two-pion correlation behavior in a small relative momentum region,
  Phys.\ Rev.\ {\bf C51} (1995) No.5, pp.R2306-R2309.

\bibitem{sinyukov}
Yu.M.Sinyukov, B.L\"{o}rstad,
 On intensity interferometry of high multiplisities,
  Z.\ Phys.\ {\bf C61} (1994) 587-592.

\bibitem{satz}
H.Satz,
  In: Proc. LHC Workshop, vol.\ 1, CERN 90-10, Geneva 1990.

\bibitem{atherton}
H.Atherton, H.Boggild, J.Boissevain et al.,
 Results from CERN experiment NA44,
  Nucl.\ Phys. {\bf A544} (1992) 125c-136c, Nos.~1,2.

\bibitem{bertsch}
G.F.Bertsch,
 Meson phase space density in heavy ion collisions from interferometry,
  Phys.\ Rev.\ Lett.\ {\bf 72} (1994) No.\ 15 pp.2349-2350. \\

\bibitem{beker}
H.Beker et al.,
 $m_{T}$ dependence of boson interferometry in heavy-ion collisions at
 the CERN SPS,
  CERN preprint CERN-PPE/94-119 (submitted to Phys.\ Rev.\ Lett.).

\end{thebibliography}
\end{document}